# Spinal Compressive Forces in Adolescent Idiopathic Scoliosis With and Without Carrying Loads: A Musculoskeletal Modeling Study

Stefan Schmid[1,2,3]*, Katelyn A. Burkhart[1,2], Brett T. Allaire[1], Daniel Grindle[1,4], Tito Bassani[5], Fabio Galbusera[5] and Dennis E. Anderson[1,2]

[1] Center for Advanced Orthopaedic Studies, Beth Israel Deaconess Medical Center, Boston, MA, United States, [2] Department of Orthopaedic Surgery, Harvard Medical School, Boston, MA, United States, [3] Spinal Movement Biomechanics Group, Division of Physiotherapy, Department of Health Professions, Bern University of Applied Sciences, Bern, Switzerland, [4] Division of Engineering Mechanics, Department of Biomedical Engineering and Mechanics, Virginia Polytechnic Institute and State University, Blacksburg, VA, United States, [5] Laboratory of Biological Structures Mechanics (LABS), IRCCS Istituto Ortopedico Galeazzi, Milan, Italy

The pathomechanisms of curve progression in adolescent idiopathic scoliosis (AIS) remain poorly understood and biomechanical data are limited. A deeper insight into spinal loading could provide valuable information toward the improvement of current treatment strategies. This work therefore aimed at using subject-specific musculoskeletal full-body models of patients with AIS to predict segmental compressive forces around the curve apex and to investigate how these forces are affected by simulated load carrying. Models were created based on spatially calibrated biplanar radiographic images from 24 patients with mild to moderate AIS and validated by comparing predictions of paravertebral muscle activity with reported values from *in vivo* studies. Spinal compressive forces were predicted during unloaded upright standing as well as standing with external loads of 10, 15, and 20% of body weight (BW) applied to the scapulae to simulate carrying a backpack in the regular way on the back as well as in front of the body and over the shoulder on the concave and convex sides of the scoliotic curve. The predicted muscle activities around the curve apex were higher on the convex side for the erector spinae (ES) and multifidi (MF) muscles, which was comparable to the EMG-based *in vivo* measurements from the literature. In terms of spinal loading, the implementation of spinal deformity resulted in a 10% increase of compressive force at the curve apex during unloaded upright standing. Apical compressive forces further increased by 50–62% for a simulated 10% BW load and by 77–94% and 103–128% for 15% and 20% BW loads, respectively. Moreover, load-dependent compressive force increases were the lowest in the regular backpack and the highest in the frontpack and convex conditions, with concave side-carrying forces in between. The predictions indicated increased segmental compressive forces during unloaded upright standing, which could be ascribed to the scoliotic deformation. When carrying loads, compressive

## INTRODUCTION

Adolescent idiopathic scoliosis (AIS) is a complex three-dimensional deformity of the spine, which affects as many as 4 out of 100 adolescents and occurs in early puberty (Cheng et al., 2015). Among congenital, neuromuscular, and other types of scoliosis, AIS is by far the most common form and is characterized by a poorly understood etiology and pathogenesis (de Seze and Cugy, 2012; Schlosser et al., 2014). Once the diagnosis of AIS is made, adolescents are generally treated conservatively using scoliosis-specific exercises (SSEs) and other forms of physiotherapy in order to minimize curve progression (Romano et al., 2012). For 10% of the initially diagnosed AIS patients, however, scoliosis exceeds a lateral deformation of 20–30° and brace treatment is indicated (Negrini et al., 2015). And finally, in about one fourth of the patients treated with exercises and braces, the progression of deformation cannot be contained and surgical intervention is required (Dolan and Weinstein, 2007).

In order to avoid bracing, surgery, and any associated health problems (e.g., chronic back pain), stopping curve progression in an early stage by means of SSEs is highly important. However, the effects of SSEs are not evident and further research is needed to clearly define the best types of SSEs as well as the frequency and intensity with which they should be administered (Romano et al., 2012). Since the pathomechanics of the AIS spine are not well understood, SSE protocols that affect spinal loading in a targeted, and scientifically sound manner cannot be developed. The literature currently lacks any studies reporting on spinal loading during functional activities or exercises in AIS patients.

Furthermore, it has been suggested that schoolbag carrying might play a role in the progression of scoliotic deformity and contribute to the development of back pain in AIS patients (Chow et al., 2006; Sahli et al., 2013). Sahli et al. (2013) recommended limiting backpack loads to 10% of body weight (BW) and carrying the load equally over both shoulders or over the shoulder on the concave rather than the convex side. However, these statements should be considered with caution since none of these studies investigated the effects of load carrying on spinal loading or trunk muscle forces.

Due to recent advancements in radiography-based geometric 3D reconstruction (Bassani et al., 2017) and musculoskeletal modeling (Bruno et al., 2015; Schmid et al., 2019), such parameters can be studies non-invasively and do not require invasive procedures such as intradiscal pressure or implant-based vertebral load measurements. The aim of this study was twofold: (1) To create subject-specific musculoskeletal full-body models of patients with mild to moderate AIS and validate predicted muscle activities with EMG data available in literature, and (2) to predict segmental compressive forces around the curve apex and how forces are affected by load carrying conditions.

## MATERIALS AND METHODS

### Development of Subject-Specific Models
#### Base Models
The base models for this study were created using our previously validated OpenSim-based musculoskeletal full-body models for children and adolescents aged 6–18 years (Schmid et al., 2019). They include a fully articulated thoracolumbar spine with a rib cage and are age- and gender-adjusted for sagittal spinal alignment as well as segmental inertial properties and maximum trunk muscle force capacity.

We enhanced the models with non-linear stiffness properties for flexion-extension and lateral-bending motions in all segments between T1/2 and L5/S1 using values from a recent meta-regression analysis over 45 studies involving experiments on adult cadaveric spines (Zhang et al., 2020). The properties were implemented using standard built-in linear bushing elements (expression-based bushing forces), which create reaction moments based on the rotational displacements of the adjacent vertebrae (Meng et al., 2015; Senteler et al., 2016). To avoid stiffness-related reaction moments in the neutral position of the spinal segments (i.e., position of the spinal segments when standing in an upright position), the bushing frames were oriented accordingly. Passive moments for segmental axial rotation were modeled using reserve coordinate actuators.

#### Implementing Spinal Deformity
Using our enhanced base models, we created subject-specific models for 24 patients with mild to moderate AIS (**Table 1**). Spinal deformity was thereby implemented based on existing simultaneously captured and spatially calibrated anterior–posterior and lateral radiographic images (EOS Imaging, France) that were acquired within a previous study conducted at the IRCCS Istituto Ortopedico Galeazzi in Milan, Italy (Bassani et al., 2017). The protocol for this study was approved by the local ethics commission and patient assent and parental permission to use the anonymized radiological data were given by signing a written informed consent form.

Three-dimensional position and orientation of each vertebra from T1 to L5 was extracted from the radiographs using a custom MATLAB script (Bassani et al., 2017). In brief, this script contains a graphical user interface (GUI), which allows for the manual identification of nine characteristic landmarks per vertebra, i.e., upper and lower vertebral corners in the sagittal and frontal planes as well as the location of spinous process in the frontal plane. Based on these landmarks, sagittal and frontal vertebral orientations were calculated as the average of the slopes of the lines connecting upper and lower vertebral corners. The axial vertebral orientation was obtained by transforming the landmark which identifies the spinous process in 3D space through evaluating a fitted referential anatomic mesh model for the vertebra under assessment (Bassani et al., 2017). To account for individual spine height, we determined the height of each vertebral body using the geometric centers of the proximal and distal intervertebral disc spaces in the sagittal plane (centroids of the lower and upper corners of the proximal and distal vertebrae, respectively) and the vertebral tilt angle in the frontal plane (**Figure 1**).

TABLE 1 | Demographics of the patients with adolescent idiopathic scoliosis (AIS) from which the biplanar radiographic images were used to create the musculoskeletal models for the current study.

| Patient | Sex | Age (years) | Height (cm) | Mass (kg) | Curve type[1] | Cobb (°) | Convexity |
|---|---|---|---|---|---|---|---|
| P1 | Female | 14 | 157 | 47 | 4B | 23.9 | Left |
| P2 | Female | 13 | 168 | 46 | 5 | 18.1 | Left |
| P3 | Female | 12 | 140 | 35 | 1B | 14.0 | Right |
| P4 | Female | 14 | 165 | 50 | 3B | 24.0 | Right |
| P5 | Female | 15 | 168 | 48 | 5 | 24.5 | Left |
| P6 | Female | 16 | 172 | 54 | 1A | 25.5 | Right |
| P7 | Female | 14 | 160 | 50 | 5 | 23.4 | Right |
| P8 | Male | 11 | 134 | 39 | 5 | 26.5 | Left |
| P9 | Female | 12 | 160 | 45 | 5 | 22.1 | Left |
| P10 | Female | 15 | 165 | 45 | 5 | 22.4 | Right |
| P11 | Male | 14 | 180 | 50 | 5 | 16.3 | Right |
| P12 | Female | 13 | 155 | 44 | 5 | 18.3 | Left |
| P13 | Male | 15 | 175 | 52 | 1A | 28.3 | Left |
| P14 | Female | 15 | 170 | 60 | 5 | 19.2 | Left |
| P15 | Female | 15 | 157 | 48 | 5 | 27.7 | Right |
| P16 | Female | 9 | 133 | 29 | 5 | 18.8 | Right |
| P17 | Female | 11 | 144 | 35 | 5 | 18.3 | Right |
| P18 | Female | 14 | 165 | 42 | 5 | 27.7 | Left |
| P19 | Female | 13 | 167 | 60 | 5 | 21.1 | Left |
| P20 | Female | 14 | 167 | 48 | 5 | 27.9 | Right |
| P21 | Female | 16 | 169 | 63 | 1A | 25.0 | Right |
| P22 | Female | 15 | 168 | 51 | 1A | 13.6 | Right |
| P23 | Female | 17 | 170 | 53 | 5 | 24.7 | Right |
| P24 | Female | 14 | 160 | 45 | 5 | 18.3 | Left |
| Average (SD) | | 13.8 (1.8) | 161.2 (12.3) | 47.5 (8.0) | – | 22.1 (4.4) | – |

[1]Curve type classification according to Lenke et al. (2001).

Subject-specific models were created in four steps: (1) scaling the base model from the corresponding age- and gender-group by body height and body mass, (2) implementing 3D spinal deformity by adjusting the orientation of the vertebral bodies T1 to L5 in the flexion/extension, lateral-flexion and axial rotation directions, (3) scaling intersegmental joint distances by vertebral body height, and (4) re-adjusting 3D orientation of the lumped head and neck segment as well as arms and ribs to a neutral position (**Figure 2**).

### Evaluation of Muscle Geometry

It was previously reported that AIS is associated with side-to-side asymmetries in erector spinae (ES) and multifidi (MF) muscle geometry (Zoabli et al., 2007; Zapata et al., 2015). To ensure appropriate handling of the muscle geometry in our models, we therefore estimated bilateral CSAs of the modeled ES and MF muscles for each thoracic and lumbar vertebral mid-plane by summing the CSAs of the individual fascicles [calculated by dividing the maximum force generating capacity of the respective fascicle by an assumed uniform maximal muscle stress (MMS) of 100 N/cm$^2$] crossing the respective mid-plane to compute an equivalent muscle group CSA at that level (Bruno et al., 2015). We refer to this procedure as a "virtual CT scan" of the model, which enables the comparison of model muscle geometry with conventional medical imaging studies. For the comparison with the literature, an asymmetry ratio was calculated by dividing the CSA of the muscles on the convex side by the CSA of the muscles on the concave side. In accordance with the *in vivo* studies, this ratio was calculated for ES at the curve apex and for MF at the levels T8, L1, and L4 as well as at the curve apex.

### Simulations

All simulations were carried out using OpenSim 3.3 (Delp et al., 2007) and MATLAB R2019a (MathWorks Inc., Natick, MA, United States). Models were solved using an inverse dynamics based static optimization with a cost function that minimized the sum of squared muscle activation (Herzog, 1987). Due to uncertainties of how the scoliotic deformation might affect muscle physiology, we solved the models without considering force-length relationships.

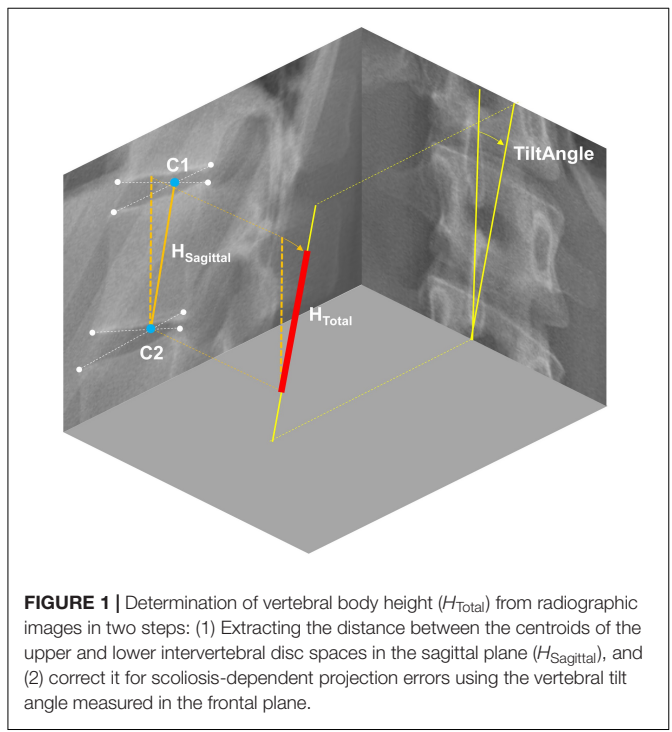

FIGURE 1 | Determination of vertebral body height ($H_{Total}$) from radiographic images in two steps: (1) Extracting the distance between the centroids of the upper and lower intervertebral disc spaces in the sagittal plane ($H_{Sagittal}$), and (2) correct it for scoliosis-dependent projection errors using the vertebral tilt angle measured in the frontal plane.

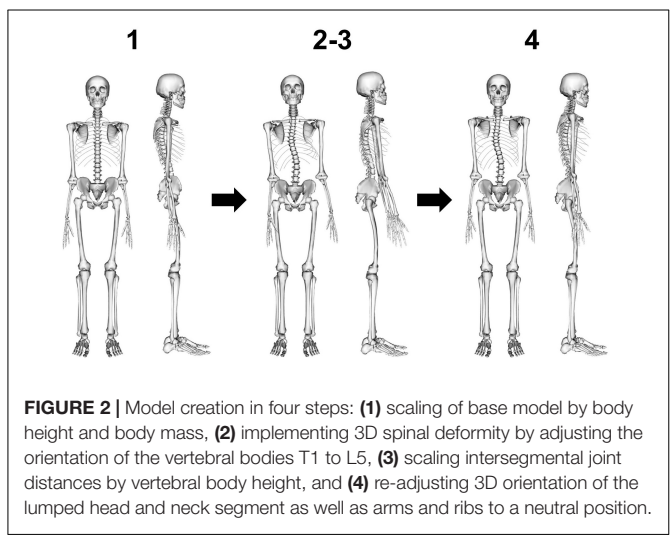

FIGURE 2 | Model creation in four steps: (1) scaling of base model by body height and body mass, (2) implementing 3D spinal deformity by adjusting the orientation of the vertebral bodies T1 to L5, (3) scaling intersegmental joint distances by vertebral body height, and (4) re-adjusting 3D orientation of the lumped head and neck segment as well as arms and ribs to a neutral position.

### Model Validation

To comply with the best practice guidelines for verification and validation of musculoskeletal models (Hicks et al., 2015), we aimed at providing a reasonable validation for our models by comparing model predictions to *in vivo* values available from the literature. Unfortunately, the literature lacks *in vivo* studies reporting on functional spinal loading (e.g., segmental compressive forces or intradiscal pressure in upright standing conditions) in AIS patients. For this reason, we conducted simulations of trunk muscle activity in standing and prone positions to compare them with the results of three reported *in vivo* studies using surface electromyography (EMG) (Cheung et al., 2005; Kwok et al., 2015; Stetkarova et al., 2016). For each of the simulations, we selected the AIS models that matched best the respective *in vivo* study population in terms of curve location.

To evaluate the accuracy of our models to predict ES muscle activity, we placed the models in a neutral upright standing position and compared the convex to concave ratios of the average activation levels of the ES muscle fascicles in the lumbar and thoracic regions as well as at the curve apex and upper and lower curve limits (two levels above and below the apex) to surface EMG-based *in vivo* measurements in AIS patients with main thoracic and thoracolumbar curves (Kwok et al., 2015) as well as patients with non-progressive AIS (Cheung et al., 2005), respectively. Accuracy of MF muscle activity predictions was evaluated by placing the models in a prone position and comparing the convex to concave ratio of the average activation levels of the MF muscle fascicles at the curve apex to needle EMG-based *in vivo* measurements in AIS patients with main thoracic curves (Stetkarova et al., 2016).

Muscle fascicles were thereby selected based on the surface electrode placement and needle electrode insertion locations described in the respective *in vivo* studies. To simulate the floor and the table where the models were "standing or lying on," we used residual point actuators with maximum activation at a force of 10 kN, which was shown to be large enough to provide the required support with minimal expenses in the static optimization (Schmid et al., 2019). Model predictions and *in vivo* measurements were compared qualitatively.

### Prediction of Compressive Forces

In order to investigate spinal compressive forces in upright standing AIS patients with and without carrying loads, we conducted simulations in five different conditions (**Table 2**).

After solving the models, joint reaction analysis was carried out to calculate the axial compressive forces acting on the spinal segments at the curve apex as well as one and two levels above and below. In addition to the absolute force magnitudes, compressive forces in AIS patients in the unloaded condition (1) were expressed as a percentage of the forces derived from the undeformed models, whereas compressive forces in AIS patients in the loaded conditions (2–5) were expressed as a percentage of the unloaded condition. To provide a coherent overview of these percentages, data were presented using violin plots with superimposed boxplots and individual values. Wilcoxon signed rank tests with an alpha-level set to 5% were conducted to test for differences from 100%.

## RESULTS

### Muscle Geometry

The predicted mean CSA ratio for the ES muscle at the apex was 1.05 SD 0.08, which compares to the ratio of 1.02 calculated

**TABLE 2** | Conditions for simulating spinal compressive forces during upright standing with and without carrying loads in patients with adolescent idiopathic scoliosis (AIS).

| Condition | Models | External load |
|---|---|---|
| Unloaded | AIS and undeformed[1] | No external load applied |
| Backpack | Only AIS | External loads of 10, 15, and 20% of body weight (BW) applied 20° dorsally angled to the lateral third of the upper edge of the scapulae (equally distributed between sides) to simulate a regular backpack |
| Frontpack | Only AIS | External loads of 10, 15, and 20% of BW applied 25° ventrally angled to the lateral third of the upper edge of the scapulae (equally distributed between sides) to simulate a backpack carried in front of the body |
| Sidepack concave | Only AIS | External loads of 10, 15, and 20% of BW applied 5° dorsally and 10° laterally angled to the lateral third of the upper edge of the scapula on the concave side to simulate a backpack carried unilaterally over the shoulder on the concave side |
| Sidepack convex | Only AIS | External loads of 10, 15, and 20% of BW applied 5° dorsally and 10° laterally angled to the lateral third of the upper edge of the scapula on the convex side to simulate a backpack carried unilaterally over the shoulder on the convex side |

*The angles for the external loads were derived from standardized photographs of one of the investigators carrying a backpack in the different modes.* [1]*Models before implementation of spinal deformity.*

from MRI-derived ES muscle volumes reported in the literature (Zoabli et al., 2007; **Figure 3**). For the MF muscle, predicted mean CSA ratios for the levels T8, L1, and L4 were 0.95 SD 0.21, 0.92 SD 0.09, and 1.01 SD 0.03, respectively. This corresponded well to the values reported in the literature for T8 (0.96), L1 (0.95), and L4 (0.98) (Zapata et al., 2015). Predicted mean CSA ratio for the MF muscle at the curve apex was 0.89 SD 0.11, however, no *in vivo* values were available for comparison.

## Validation Studies

The predicted mean ES muscle activity ratios during upright standing indicated higher activity on the convex side of the muscle for thoracic (thoracic region: 1.29 SD 0.71; lumbar region: 1.18 SD 0.78) and thoracolumbar curves (thoracic region: 1.05 SD 0.19; lumbar region: 1.12 SD 0.31) (**Figure 4**). Furthermore, ES muscle activity within the curve also indicated higher convex activity at the apex (1.72 SD 1.06) as well as the upper (1.37 SD 0.78) and lower curve ends (1.02 SD 0.83). These ratios compare reasonably well to the literature for the thoracic and lumbar portions of the ES muscle in patients with main thoracic curves (1.87 SD 1.7 and 1.7 SD 0.85, respectively) as well as at the curve apex (2.1 SD 1.38) and the lower curve end (0.96 SD 0.32) (Cheung et al., 2005; Kwok et al., 2015). For MF muscle activity, models predictions indicated higher activity at the apex on the convex side of the curve (1.21 SD 0.50), which goes along with the *in vivo* measured ratio of 1.38 (Stetkarova et al., 2016).

## Spinal Compressive Forces

The implementation of spinal deformity resulted in higher median compressive forces within the scoliotic curve, with forces constantly increasing from two levels above the apex (103% IQR 13%, $p = 0.092$), to the apex (110% IQR 13%, $p < 0.001$), to two levels below the apex (118% IQR 16%, $p < 0.001$) (**Figure 5**).

All loaded median axial compressive forces were significantly different from 100% (i.e., from an unloaded condition) at a level of $p < 0.001$. When carrying a load which corresponded to 10% of BW, median axial compressive forces increased on average by 50% for the regular backpack, 62% for the backpack carried in front and 60% and 54% for the backpack carried over the shoulder on the convex and concave sides, respectively (**Figure 6**). In the regular backpack as well as both sidepack conditions, compressive force increased the most above the apex and the least below the apex, whereas in the frontpack condition, compressive force increased about equally on all spinal levels.

When applying loads corresponding to 15 and 20% of BW, median axial compressive forces increased on average by 77 and 103% for the backpack, 93 and 125% for the frontpack, 94 and 128% for the convex sidepack, and 85 and 116% for the concave sidepack conditions, respectively (**Figures 7, 8**). The load distribution pattern within the spinal levels of the scoliotic curve remained similar as described for the 10% of BW load.

A complete set of the absolute force magnitude and relative force percentage values can be found in the **Supplementary Material**.

## DISCUSSION

We created 24 subject-specific musculoskeletal full-body models from biplanar radiographic images of patients with mild to moderate AIS and validated these models by comparing predictions of paravertebral muscle activity with reported values from *in vivo* studies. Moreover, we predicted apical spinal loads with and without simulated load carrying, i.e., carrying a backpack in the regular way, carrying a backpack in front of the body and carrying a backpack over the shoulder on the concave and convex sides of the scoliotic curve.

The evaluation of muscle geometry indicated that the implementation of spinal deformity resulted in side-to-side asymmetries, which agreed with reports in the literature. The validation studies showed higher convex ES and MF muscle activity around the curve apex, which was comparable to the EMG-based *in vivo* measurements from the literature. Measurements of overall thoracic and lumbar ES muscle activity agreed well for thoracic but less thoracolumbar curves. In terms of spinal loading, the implementation of spinal deformity resulted in a 10% increase of compressive force at the curve apex during unloaded upright standing. Apical compressive forces further increased by 50–62% for a simulated 10% BW load and by 77–94% and 103–128% for 15 and 20% BW loads, respectively. Moreover, load-dependent compressive force increases were the lowest in the regular backpack and the highest in the frontpack and convex conditions, with concave side-carrying forces in between.

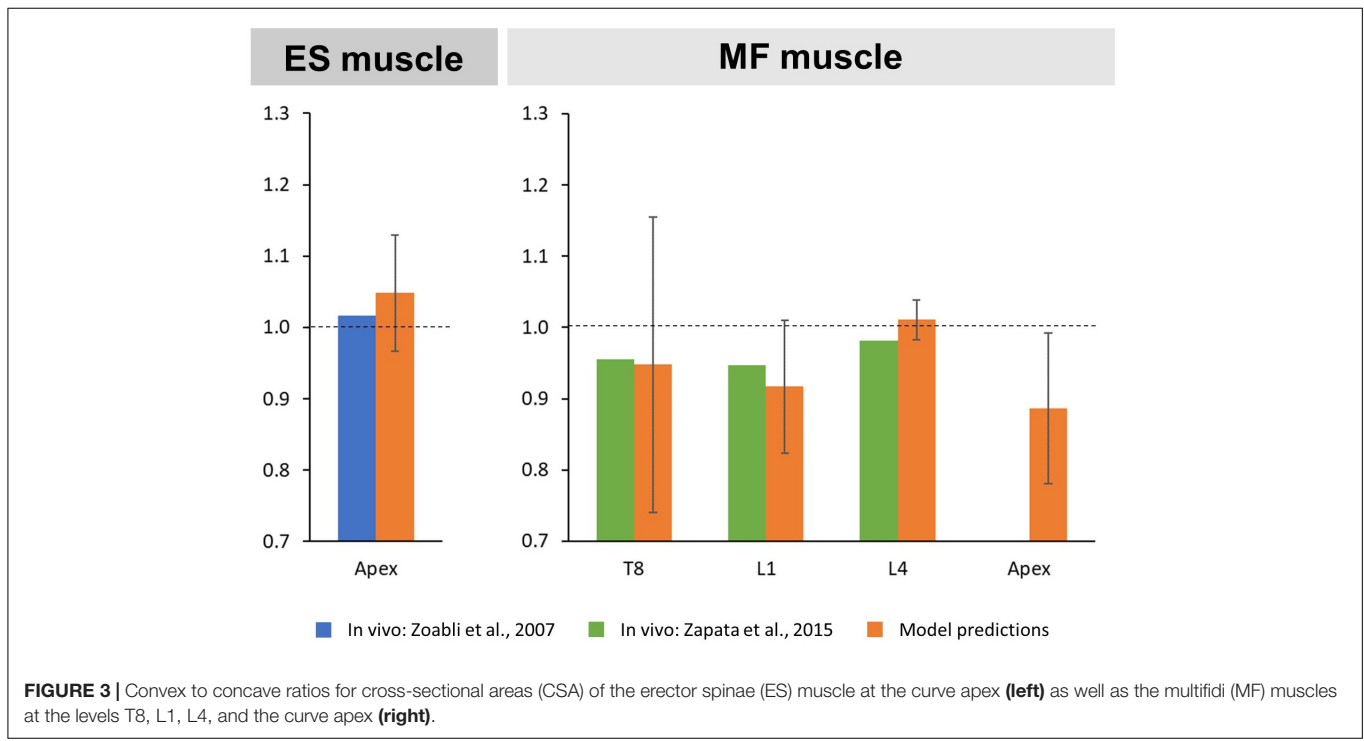

**FIGURE 3 |** Convex to concave ratios for cross-sectional areas (CSA) of the erector spinae (ES) muscle at the curve apex **(left)** as well as the multifidi (MF) muscles at the levels T8, L1, L4, and the curve apex **(right)**.

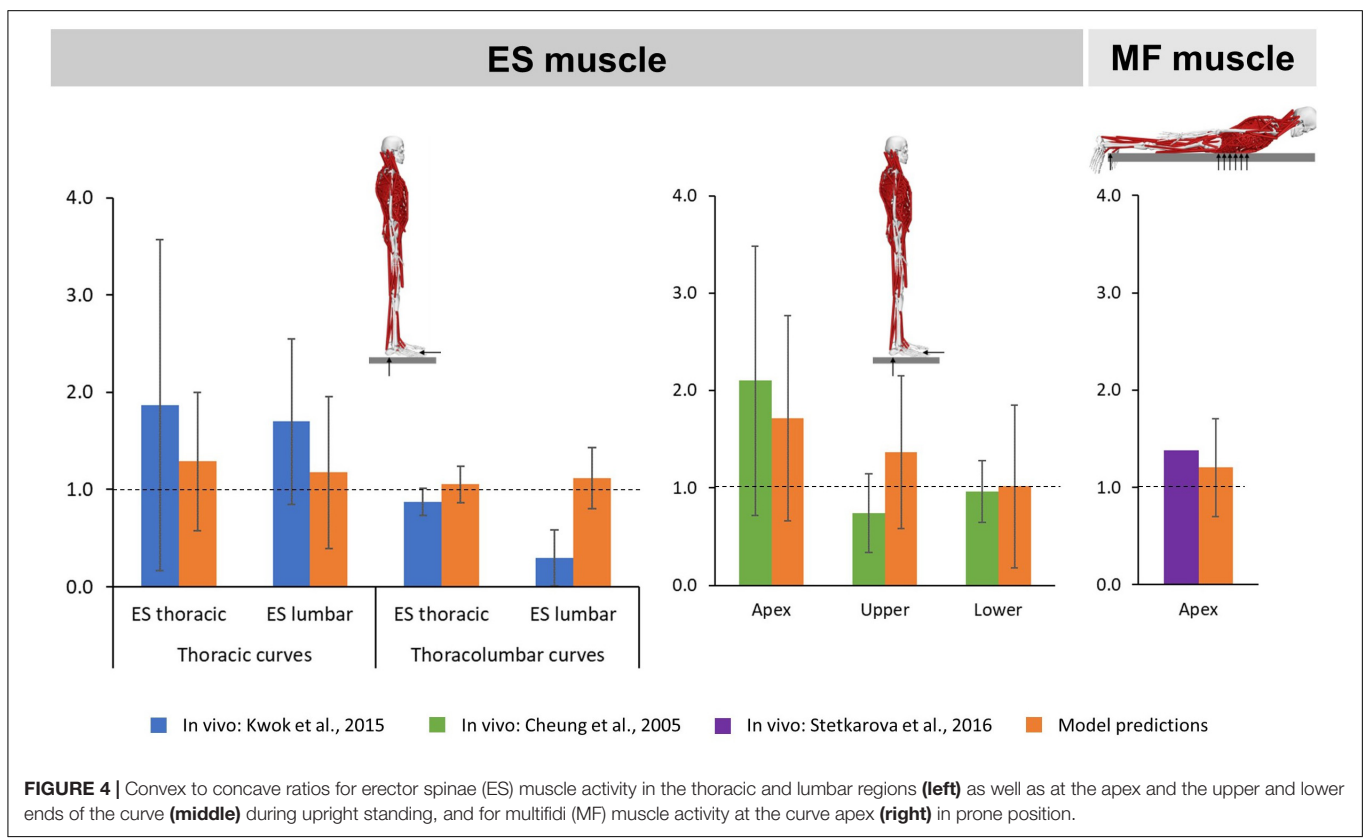

**FIGURE 4 |** Convex to concave ratios for erector spinae (ES) muscle activity in the thoracic and lumbar regions **(left)** as well as at the apex and the upper and lower ends of the curve **(middle)** during upright standing, and for multifidi (MF) muscle activity at the curve apex **(right)** in prone position.

Even though the evaluation of muscle geometry estimated from our models indicated larger ES muscle CSA on the convex side of the scoliotic deformation, about one third of our models showed in fact larger ES muscle CSA on the concave side. This agrees with the study of Zoabli et al. (2007), reporting that even though there was an overall tendency for larger ES muscle volume on the convex side, some patients also presented larger volumes on the concave side. For the MF muscle at the curve apex, a larger CSA on the concave side was found in about 90% of the models, whereas the remaining 10% showed a larger CSA on the convex side. Despite these uncertainties, however, the muscle geometry estimated from our AIS models was comparable to what was measured *in vivo*. This raises the question of whether the previously reported muscle volume/thickness asymmetries in AIS patients represent an actual change in muscle, or is just a result of measurement of muscle with different spinal geometry.

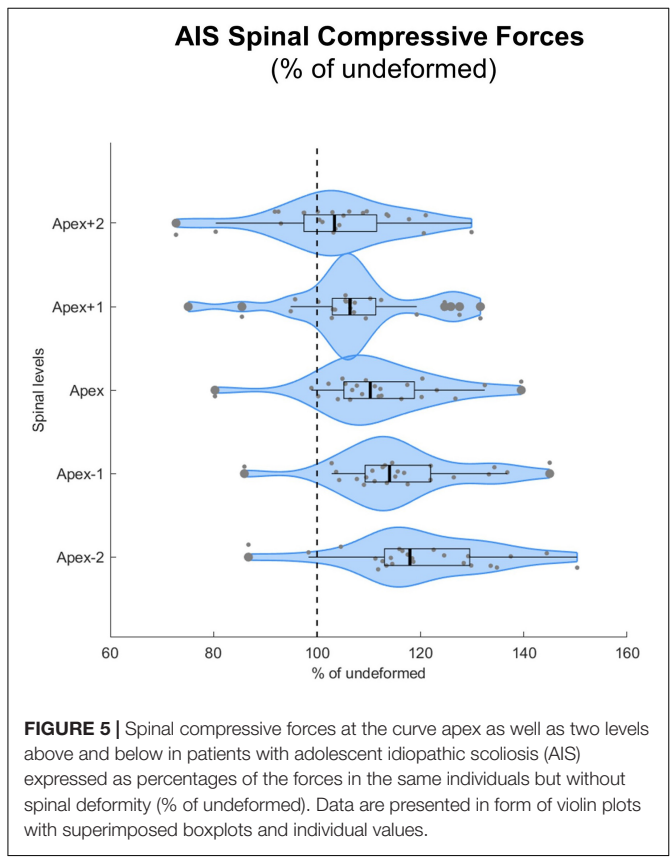

**FIGURE 5 |** Spinal compressive forces at the curve apex as well as two levels above and below in patients with adolescent idiopathic scoliosis (AIS) expressed as percentages of the forces in the same individuals but without spinal deformity (% of undeformed). Data are presented in form of violin plots with superimposed boxplots and individual values.

The partial disagreements between our model predictions and the ES muscle activity ratios derived from the results reported by Kwok et al. (2015) might be related to differences in curve characteristics (i.e., location of the apex and severity of deformation) between the respective patient populations. Especially when evaluating overall ES muscle activity in the thoracic and lumbar regions with the same electrode placement for all patients, different curve characteristics could have a significant effect on muscle activation. Furthermore, muscle fiber redistribution with a higher proportion of type I fibers on the convex side of the AIS curve (Gonyea et al., 1985; Meier et al., 1997; Mannion et al., 1998; Stetkarova et al., 2016) might be another contributing factor, since muscle fiber type seems to have an influence on the EMG signal (Poosapadi Arjunan et al., 2016). However, due to the lack of appropriate data, the consideration of fiber distribution change in our current models would be associated with too many assumptions. Finally, it is not known whether AIS has an influence on the force-length-relationship of the paravertebral muscles, which could also have an influence on EMG activity. Based on our validation studies, however, we consider the muscle activation patterns predicted by our models comparable to the patterns reported in the *in vivo* studies.

This is the first study using inverse dynamics-based musculoskeletal full-body modeling to investigate the immediate effect of AIS-related spinal deformity on axial compressive forces within the scoliotic curve. The results suggested that spinal deformity causes an overall increase in compressive forces, with forces increasing the most at the lower and the least at the upper end of the curve. When considering the different curve types, it appears that the average increase in segmental loading was not significantly related to the spinal level of the curve apex or the curve severity, i.e., the Cobb angles (Pearson correlation: $r = 0.28$, $p = 0.193$ and $r = 0.22$, $p = 0.302$, respectively). It should be considered, however, that all the patients in this study had mild to moderate AIS, providing a relatively small range of Cobb angles. The relationship between compressive force increase and curve severity would probably be more pronounced with larger range of Cobb angles in the dataset. It should also be noted that a small fraction of the predictions resulted in compressive force decreases. When looking at these cases, especially those predicting compressive forces of less than 90% of the undeformed, it appears that these patients tended to have particularly flat thoracic sagittal profiles (i.e., <25° of thoracic kyphosis), which might have resulted in a significant reduction of muscular effort (specifically the ES muscle) and therefore in reduced compressive forces.

The simulation of load carrying in AIS patients indicated compressive force increments that were dependent on the carrying mode as well as the weight of the load. Carrying the load in front of the body resulted in considerably higher compressive forces compared to regular backpack carrying. This is not surprising since the front carrying mode would be assumed to cause higher muscular effort to prevent increased flexion. Interestingly, carrying the backpack on the concave side yielded compressive forces that were slightly higher than the ones for regular backpack carrying, but lower compared to carrying the load in front. This indicates that shifting the load from the back to the concave side does result in increased compressive forces – most likely due to the higher muscular effort on the contralateral side – but not as much as carrying the load symmetrically in front. In addition, carrying the load on the convex side caused compressive forces that were comparable to the front carrying mode, most likely because the load was acting more directly on the spinal segments. The different patterns of compressive force increments between the frontpack condition (constant over all levels) and the other carrying conditions (decrease in force increments from two levels above to two levels below the apex) might be related to the direction of the applied external load in the sagittal plane. In the frontpack condition, the external load was directed anteriorly, which resulted in increased paraspinal muscle activity. In the other carrying conditions, however, the external load is directed posteriorly, causing an increased activity predominantly of the abdominal muscles. Considering that most of the abdominal muscles are only indirectly connected to the spine, i.e., over the rib cage, it seems plausible that this affected spinal loading differently than in the frontpack condition. In any way, these results do not allow any conclusions on whether carrying a load in front, on either side or regularly on the back is advantageous to minimize curve progression or prevent complications such as joint degeneration or back pain. In fact, it is possible that carrying the load on the convex side might put the patients at a higher risk for back pain, but at the same time slow down curve progression by modulating the segmental load in a way that vertebral growth is positively affected due to the Hueter-Volkmann law, where bone growth is slowed with compression and accelerated with distraction (de Seze and Cugy, 2012). Future studies should therefore address this issue by using a combination of motion capture-driven musculoskeletal and finite element models.

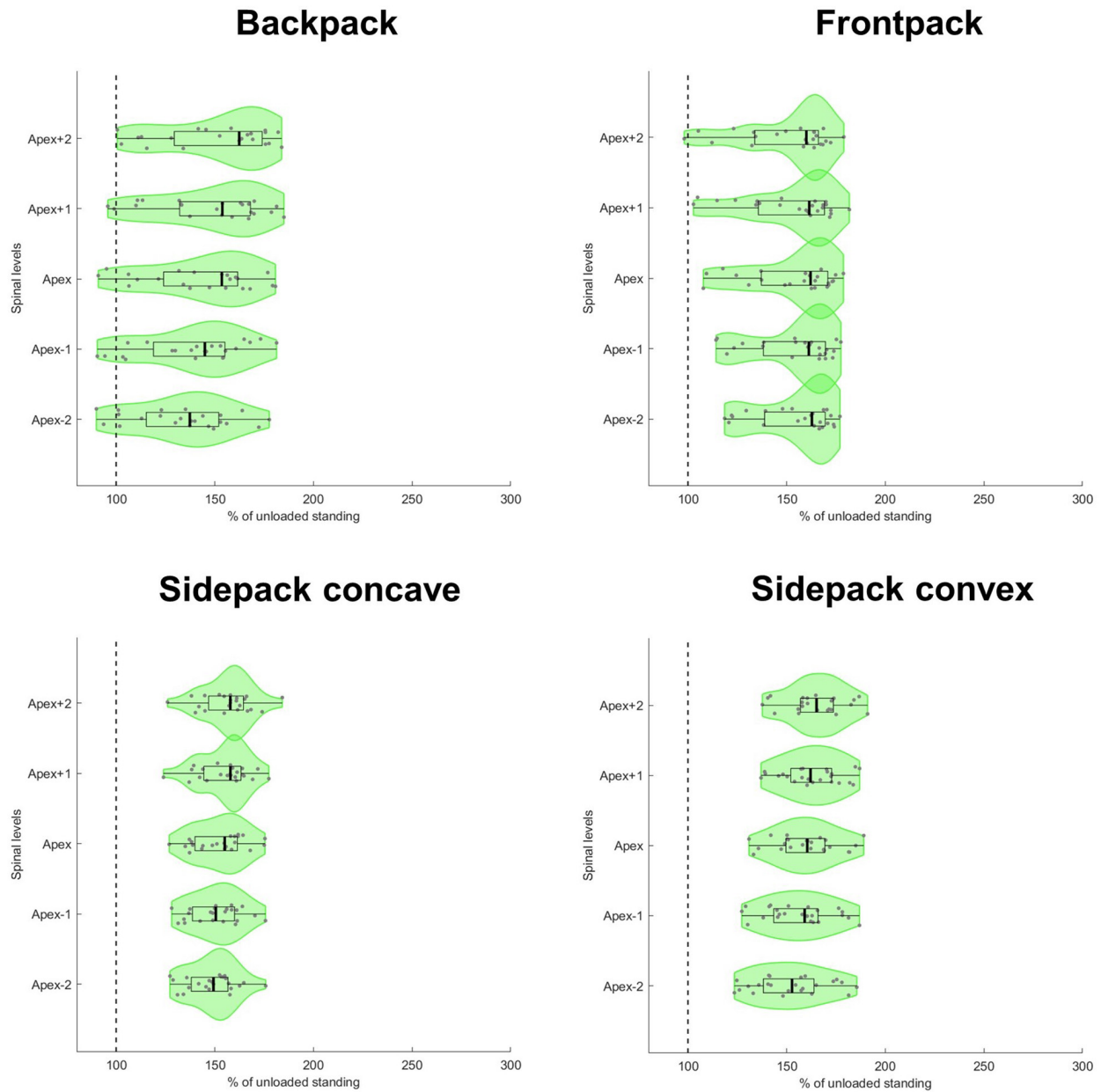

**FIGURE 6** | Spinal compressive forces at the curve apex as well as two levels above and below in patients with adolescent idiopathic scoliosis (AIS) when carrying a backpack with a weight corresponding to 10% of body weight in the regular way **(top left)**, in front of the body **(top right)** as well as over the shoulder on the concave **(bottom left)**, and convex sides **(bottom right)** of the scoliotic curve. Forces are expressed as percentages of unloaded upright standing and presented in form of violin plots with superimposed boxplots and individual values.

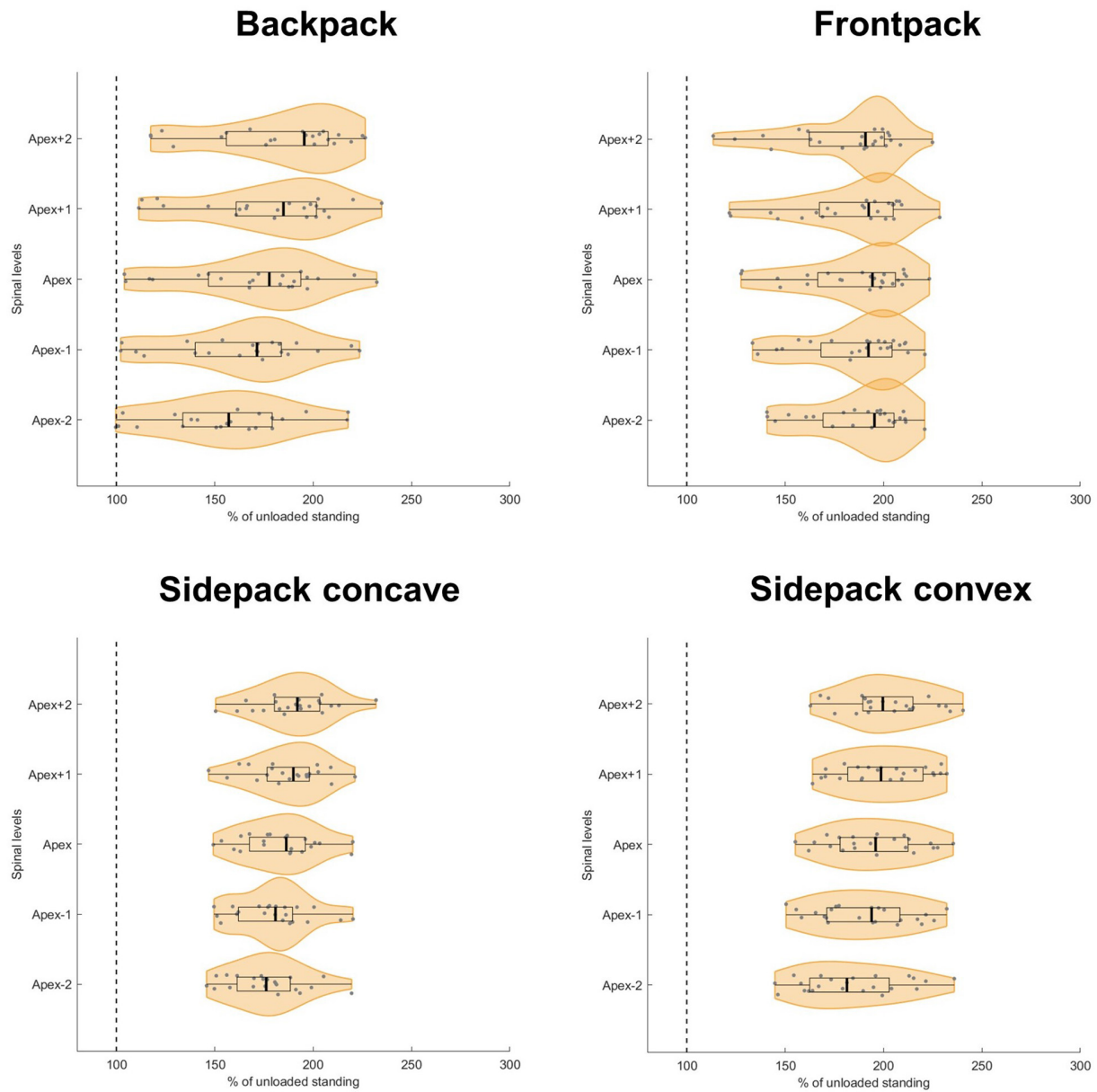

**FIGURE 7** | Spinal compressive forces at the curve apex as well as two levels above and below in patients with adolescent idiopathic scoliosis (AIS) when carrying a backpack with a weight corresponding to 15% of body weight in the regular way **(top left)**, in front of the body **(top right)** as well as over the shoulder on the concave **(bottom left)**, and convex sides **(bottom right)** of the scoliotic curve. Forces are expressed as percentages of unloaded upright standing and presented in form of violin plots with superimposed boxplots and individual values.

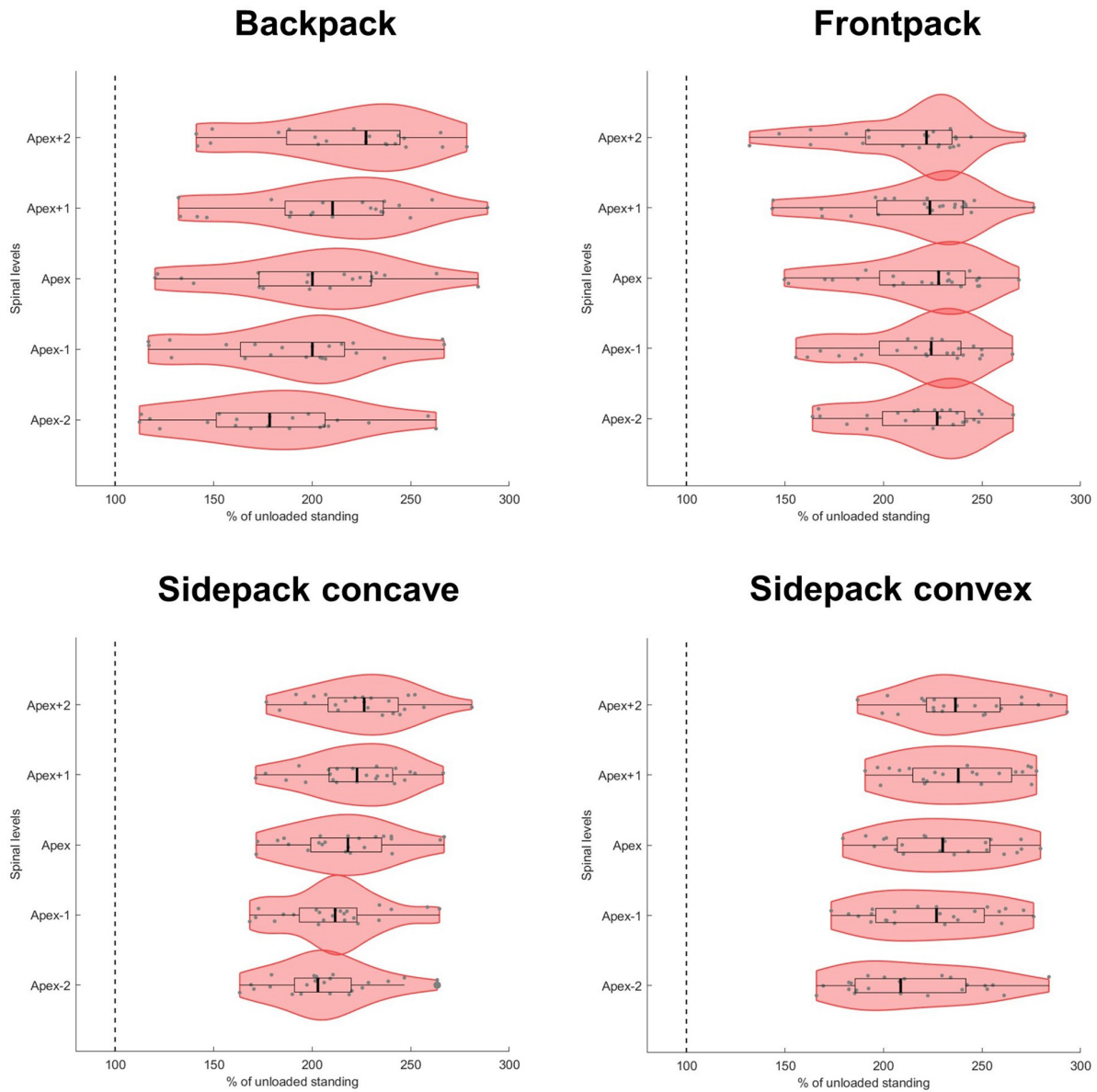

FIGURE 8 | Spinal compressive forces at the curve apex as well as two levels above and below in patients with adolescent idiopathic scoliosis (AIS) when carrying a backpack with a weight corresponding to 20% of body weight in the regular way **(top left)**, in front of the body **(top right)** as well as over the shoulder on the concave **(bottom left)**, and convex sides **(bottom right)** of the scoliotic curve. Forces are expressed as percentages of unloaded upright standing and presented in form of violin plots with superimposed boxplots and individual values.

This study has some important limitations that should be discussed. First of all, due to a lack of appropriate data for healthy children and adolescents as well as patients with AIS, the passive segmental stiffness properties that were implemented in the current base models were derived from healthy adult cadaveric spines. This issue could be addressed in the future by conducting clinical studies in AIS patients using approaches such as the intraoperative determination of load-displacement behavior proposed by Reutlinger et al. (2012). However, the lack of appropriate stiffness properties did not affect the current predictions since all simulations were conducted in a neutral position, i.e., with the spinal segments assumed to be in the neutral zone (Smit et al., 2011). Stiffness-related reaction moments would only have occurred with induced segmental rotations. When using the models for future investigations involving simulations beyond the neutral position of the spine, on the other hand, these limitations will have to be considered. Secondly, we did only consider compressive forces in this study, but forces in other directions (i.e., anterior–posterior and medial–lateral shear forces) might also be strongly affected by scoliotic deformities. Furthermore, the patient populations of the *in vivo* studies used for the validation of our models did not exactly match the population from which they were created. It is therefore advised that future studies investigating spinal loading in AIS include EMG measurements of the paraspinal muscles for more specific validations of the respective models. Lastly, the current simulations were not based on real-life kinematics, i.e., they were not driven by motion capture data. Especially for the load carrying investigations, it can be assumed that real subjects would have slightly adapted their posture based on the applied load, such as previously observed for regular backpack carrying in healthy young adults (Neuschwander et al., 2010).

In conclusion, this study used validated subject-specific OpenSim-based musculoskeletal full-body models to provide an insight into spinal loading in patients with AIS with and without carrying loads. The predictions indicated increased segmental compressive forces of about 10% around the curve apex during unloaded upright standing. When carrying loads, compressive forces further increased depending on the carrying mode and the weight of the load. These results can be used as a basis for further studies investigating segmental loading in AIS patients during functional activities. Models can thereby be created using the same approach as proposed in this study.

## DATA AVAILABILITY STATEMENT

The datasets generated for this study are available on request to the corresponding author.

## AUTHOR CONTRIBUTIONS

## FUNDING

This work was funded by the Swiss National Science Foundation (SNSF, grant no. 178427) and the National Center for Simulation in Rehabilitation Research (NCSRR, sub-award of NIH grant no. 5P2CHD065690).

## ACKNOWLEDGMENTS

The authors thank Dr. Bram Verhofste for assistance in the clinical interpretation of the radiographs.

## SUPPLEMENTARY MATERIAL

The Supplementary Material for this article can be found online at: https://www.frontiersin.org/articles/10.3389/fbioe.2020.00159/full#supplementary-material